\documentclass[fleqn,usenatbib]{mnras}

\usepackage{newtxtext,newtxmath}

\usepackage[T1]{fontenc}

\DeclareRobustCommand{\VAN}[3]{#2}
\let\VANthebibliography\thebibliography
\def\thebibliography{\DeclareRobustCommand{\VAN}[3]{##3}\VANthebibliography}

\usepackage{graphicx}	
\usepackage{amsmath}	
\usepackage[dvipsnames]{xcolor}

\title[Characterizing dense cores using HROs]{The JCMT BISTRO Survey: The magnetised evolution of star-forming cores in the Ophiuchus Molecular Cloud interpreted using Histograms of Relative Orientation}

\author[J. P. Perry et al.]{
James P. Perry,$^{1}$\thanks{E-mail: james.perry.21@ucl.ac.uk} Kate Pattle,$^{1}$ Doug Johnstone,$^{2}$ Woojin Kwon,$^{3,4}$ Tyler Bourke,$^{5,6}$ Eun Jung Chung,$^{7}$\newauthor Simon Coud\'e,$^{8,9}$ Yasuo Doi,$^{10}$ Lapo Fanciullo,$^{11}$ Jihye Hwang,$^{12}$ Zacariyya A. Khan,$^{1}$ Jungmi Kwon,$^{13}$ \newauthor Shih-Ping Lai,$^{14,15}$ Valentin J. M. Le Gouellec,$^{16}$ Chang Won Lee,$^{12,17}$ Nagayoshi Ohashi,$^{15}$ \newauthor Sarah Sadavoy,$^{18}$ Giorgio Savini,$^{1}$ Ekta Sharma,$^{19}$ Motohide Tamura$^{20,13,21}$
\\
$^{1}$Department of Physics and Astronomy, University College London, WC1E 6BT London, UK\\
$^{2}$NRC Herzberg Astronomy and Astrophysics, 5071 West Saanich Road, Victoria, BC V9E 2E7, Canada\\
$^{3}$Department of Earth Science Education, Seoul National University, 1 Gwanak-ro, Gwanak-gu, Seoul 08826, Republic of Korea\\
$^{4}$SNU Astronomy Research Center, Seoul National University, 1 Gwanak-ro, Gwanak-gu, Seoul 08826, Republic of Korea\\
$^{5}$SKA Observatory, Jodrell Bank, Lower Withington, Macclesfield SK11 9FT, UK\\
$^{6}$Jodrell Bank Centre for Astrophysics, School of Physics and Astronomy, University of Manchester, Oxford Road, Manchester, M13 9PL, UK\\
$^{7}$Department of Astronomy and Space Science, Chungnam National University, 99 Daehak-ro, Yuseong-gu, Daejeon 34134, Republic of Korea\\
$^{8}$Department of Earth, Environment, and Physics, Worcester State University, Worcester, MA 01602, USA \\
$^{9}$Center for Astrophysics | Harvard \& Smithsonian, 60 Garden Street, Cambridge, MA 02138, USA \\
$^{10}$Department of Earth Science and Astronomy, Graduate School of Arts and Sciences, The University of Tokyo, 3-8-1 Komaba, Meguro, \\ Tokyo 153-8902, Japan\\
$^{11}$National Chung Hsing University, 145 Xingda Rd., South Dist., Taichung City 402, Taiwan\\
$^{12}$Korea Astronomy and Space Science Institute, 776 Daedeokdae-ro, Yuseong-gu, Daejeon 34055, Republic of Korea\\
$^{13}$Department of Astronomy, Graduate School of Science, The University of Tokyo, 7-3-1 Hongo, Bunkyo-ku, Tokyo 113-0033, Japan\\
$^{14}$Institute of Astronomy and Department of Physics, National Tsing Hua University, Hsinchu 30013, Taiwan\\
$^{15}$Academia Sinica Institute of Astronomy and Astrophysics, No.1, Sec. 4., Roosevelt Road, Taipei 10617, Taiwan\\
$^{16}$Universit\'e Paris-Saclay, CNRS, CEA, Astrophysique, Instrumentation et Mod\'elisation de Paris-Saclay, 91191 Gif-sur-Yvette, France\\
$^{17}$University of Science and Technology, Korea, 217 Gajeong-ro, Yuseong-gu, Daejeon 34113, Republic of Korea\\
$^{18}$Department for Physics, Engineering Physics and Astrophysics, Queen's University, Kingston, ON, K7L 3N6, Canada\\
$^{19}$CAS Key Laboratory of FAST, National Astronomical Observatories, Chinese Academy of Sciences, People's Republic of China\\
$^{20}$National Astronomical Observatory of Japan, National Institutes of Natural Sciences, Osawa, Mitaka, Tokyo 181-8588, Japan\\
$^{21}$Astrobiology Center, National Institutes of Natural Sciences, 2-21-1 Osawa, Mitaka, Tokyo 181-8588, Japan
}

\date{Accepted XXX. Received YYY; in original form ZZZ}

\pubyear{2024}

\begin{document}
\label{firstpage}
\pagerange{\pageref{firstpage}--\pageref{lastpage}}
\maketitle

\begin{abstract}
{The relationship between B-field orientation and density structure in molecular clouds is often assessed using the Histogram of Relative Orientations (HRO). {We perform a plane-of-the-sky geometrical analysis of projected B-fields}, by interpreting HROs in dense, spheroidal, prestellar and protostellar cores. We use James Clerk Maxwell Telescope (JCMT) POL-2 850 $\mu$m polarisation maps and \textit{Herschel} column density maps to study dense cores in the Ophiuchus molecular cloud complex. We construct two-dimensional core models, assuming Plummer column density profiles and modelling both linear and hourglass B-fields. We find high-aspect-ratio ellipsoidal cores produce strong HRO signals, as measured using the shape parameter $\xi$. Cores with linear fields oriented $< 45^{\circ}$ from their minor axis produce constant HROs with $-1 < \xi < 0$, indicating fields are preferentially parallel to column density gradients. Fields parallel to the core minor axis produce the most negative value of $\xi$. For low-aspect-ratio cores, $\xi \approx 0$ for linear fields. {Hourglass fields produce a minimum in $\xi$ at intermediate densities in all cases, converging to the minor-axis-parallel linear field value at high and low column densities.} We create HROs for six dense cores in Ophiuchus. $\rho$ Oph A and IRAS 16293 have high aspect ratios and preferentially negative HROs, consistent with moderately strong-field behaviour. $\rho$ Oph C, L1689A and L1689B have low aspect ratios, and $\xi \approx 0$. $\rho$ Oph B is too complex to be modelled using a simple spheroidal field geometry. We see no signature of hourglass fields, agreeing with previous findings that dense cores generally exhibit linear fields on these size scales.}
\end{abstract}



\section{Introduction}
Magnetic fields permeate the interstellar medium (ISM), yet their role in star formation is not yet well-understood. {Magnetic fields are thought to be acting as a guide to collapsing} elements by confining the motion of particles and providing a direction and structure along which collapse can propagate \citep{hennebelle2013}. They are also believed to inhibit protostar formation by suppressing dense gas formation \citep{ntormousi2017} and slowing the growth of perturbative instabilities \citep{Fiege_Pudritz_2000a,Fiege_Pudritz_2000b}. Yet their importance in controlling the rate of star formation and their relative importance to turbulence and gravity is still debated \citep{Hennebelle_Chabrier_2011,Krumholz_Federrath_2019}.

Of interest is the magnetic field within prestellar and protostellar cores. A prestellar core is a gravitationally bound dense core \citep{Ward-Thompson_et_al_1994}, likely formed from the fragmentation of a magnetically supercritical (gravitationally dominated) filament \citep{zhang2022}, which has a high probability of forming a star \citep{Konyves_et_al_2015}. Its density is high and continues to rise {over time}, but there is no protostar yet formed at its centre. Since these cores are starless, they are the immediate predecessors of the Class 0 protostar phase and are therefore the initial conditions associated with protostar formation \citep{Andre_Montmerle_1994, Ward-Thompson_et_al_1994}.  A protostellar core is a dense star-forming core with one or more hydrostatic objects -- i.e. envelope-dominated Class 0 protostar(s) \citep{andre1993} -- embedded within it.

Prestellar core systems with a high initial degree of magnetisation are theorised to have an `hourglass' magnetic field structure \citep{Mouschovias_1976,myers_basu_2021}. In this scenario, gas collects along the field lines until self-gravity dominates and pulls both the gas and the field inward.  However, observations of prestellar cores have as yet not yielded a definitive detection of such a magnetic field \citep{pattle_et_al_2023}. Instead, prestellar cores are often observed to show linear fields with a relative angle of less than $\sim30$ deg to {their} minor axis \citep{basu_2000,ward_thompson_et_al_2000}.

Hourglass magnetic fields have been observed in a range of protostellar cores, including low-mass cores \citep{Girart_et_al_2006,maury2018}, high-mass cores \citep{qiu_et_al_2014}, and protobinary cores \citep{Kwon_et_al_2019}.  To date, all observations of hourglass fields in protostellar cores have been made with interferometers.

{Polarimetry of thermal dust emission from dust grains aligned with the magnetic field} is one of the main techniques used to study the alignments of magnetic fields in both star-forming filaments and cores. A linear polarimeter can map the preferential polarisation of {dust grains aligned with the magnetic field in} the interstellar medium, which is then {used to infer the alignment of the magnetic field orientation with respect to the underlying physical density structure (e.g., molecular clouds, star-forming regions).} But since estimating magnetic field strengths in dense cores remains difficult (e.g. \citealt{pattle_et_al_2023}), comparison between data and models is key to understanding the dynamic importance of magnetic fields in the ISM. A useful tool for this is the histogram of relative orientations (HRO, \citealt{Soler_et_al_2013}), which compares the relative orientation of column density gradients and magnetic field lines. {Column densities are determined by measuring thermal dust emission. Emissions are used to determine a dust opacity at a given observation wavelength, which are then converted into a visual extinction magnitude, $A_{\text{V}}$. This value is then converted to hydrogen column density, $N_{\text{H}}$, via $ N_{\text{H}} / A_{\text{V}} = 1.8 \times 10^{21}$ cm$^{-2}$ \citep{Gratier_et_al_2021}}.
It has been seen in many instances that on filament scales there are transitions from parallel relative orientation to perpendicular as gravity comes to dominate \citep[see e.g.][]{Planck_Collaboration_2016,Kusune_et_al_2019,Pillai_et_al_2020,Lee_et_al_2021,Kwon_et_al_2022}. However, the results for HROs in prestellar cores are not well studied. 

{We note there are alternative techniques available for assessing the relationship between magnetic field and density structure in clouds, for example, the Hessian technique \citep[e.g.][]{Polychroni_et_al_2013} and the Rolling Hough Transform (RHT, \citealt{Clark_et_al_2014}). However, since HROs are a very widely used measure, \citep[e.g.][]{Planck_Collaboration_2016,soler_et_al_2017,Lee_et_al_2021,Wang_et_al_2024}, including for use in interpreting magnetic fields in dense cores \citep{Kwon_et_al_2022}, we seek to assess the interpretation of this measure in the environment of dense cores. In this work therefore, we focus on the HRO, while noting there are other techniques available.}

In this work, we aim to investigate the expected behaviour of HROs in dense, spheroidal cores, and so to provide a means by which to interpret the relationship between column density and magnetic field structure in prestellar and protostellar cores without recourse to detailed modelling. {We perform a geometrical analysis of the projected B-field on the plane of the sky (POS). No information about the 3D structure of density and B-field was used in this work.} To do so, we construct simple models, and compare these to the HROs of dense clumps and cores in the nearby Ophiuchus molecular cloud. We present HROs for Ophiuchus and its subsets, the Lynds dark clouds L1688 and L1689 \citep{Lynds_1962}, using magnetic field data from the POL-2 polarimeter on the James Clerk Maxwell Telescope (JCMT), observed as part of the B-fields in Star-Forming Region Observations (BISTRO) Survey \citep{wardthompson2017}. We also use column density maps derived from \textit{Herschel} Space Observatory data as part of the \textit{Herschel} Gould Belt Survey \citep{andre2010}.  We study in detail three regions within L1688 ($\rho$ Oph A, B and C) and three within L1689 (IRAS 16293-2422, hereafter referred to as IRAS 16293, along with L1689A and L1689B). Using the results of fitting two-dimensional Gaussians to the column density maps, we construct toy models, comparing the resultant HRO to the observed data. 

In Section~\ref{sec:observations} we describe the observational data. In Section~\ref{sec:results} we describe the HRO technique, the model setup and analysis performed, before presenting the results. We interpret our results in Section~\ref{sec:discussion}, then summarise and conclude in Section~\ref{sec:conclusion}.

\section{Observations} \label{sec:observations}

\subsection{James Clerk Maxwell Telescope}

We inferred magnetic field {orientation on the plane of the sky} for the dense cores in Ophiuchus using 850 $\mu$m dust polarisation observations made using the POL-2 polarimeter \citep{friberg2016} on the SCUBA-2 camera \citep{holland2013} on the JCMT. {The polarisation maps have an angular resolution of $14.1^{\prime\prime}$ \citep{Soam_et_al_2018}.}

The data on $\rho$ Oph A, B and C and L1689B were taken as part of the JCMT BISTRO Survey, under project code M16AL004.  The data on IRAS 16293-2422 and L1689A were taken under project code M19AP038.  $\rho$ Oph A was first published by \citet{Kwon_et_al_2018}, $\rho$ Oph B by \citet{Soam_et_al_2018}, and $\rho$ Oph C by \citet{liu2019}. The data used in this work were presented by \citet{pattle2019} ($\rho$ Oph A, B and C), and by \citet{Pattle_et_al_2021} (IRAS 16293, L1689A and L1689B).  Please see those works for a detailed description of the observations and the POL-2 data reduction process.

\subsection{\textit{Herschel} Space Observatory}

The $N({\rm H}_{2})$ column density map used in this work was presented by \citet{ladjelate2020}, and is the result of spectral energy distribution (SED) fitting to data taken using the PACS and SPIRE cameras on the \textit{Herschel} Space Observatory as part of the \textit{Herschel} Gould Belt Survey \citep{andre2010}. {The SED fitting performed by \citet{ladjelate2020} used the observed $36.3^{\prime\prime}$ 500 $\mu$m map to estimate a column density map at the SPIRE 250 $\mu$m resolution of $18.2^{\prime \prime}$. We used the $18.2^{\prime \prime}$ high resolution column density map of Ophiuchus \citep[for details see][]{ladjelate2020}}. The column density map is available at \url{http://www.Herschel.fr/cea/gouldbelt/en/}. 

\subsection{{Matching resolutions}}
{We smoothed the $14.1^{\prime \prime}$ resolution JCMT POL-2 data to a resolution of $18.2^{\prime \prime}$ using a Gaussian filter prior to analysis, matching the high resolution \textit{Herschel} column density maps. The \textit{Herschel} maps were then re-projected onto the pixel grid of the smoothed JCMT data. }

\subsection{Magnetic field and column density angles} \label{sec: theta_phi_definitions}
The {observed} polarisation angle, $\theta^{\prime}$, of {dust emission is given by}
\begin{equation}\label{eq:polarisation_angle}
    \theta^{\prime} = \frac{1}{2} \text{arctan}\left(\frac{U}{Q}\right),
\end{equation}
where $Q$ and $U$ are the Stokes $Q$ and $U$ parameters determined from the JCMT polarisation maps {of dust alignment. Asymmetric dust grains in molecular clouds tend to align themselves with their major axis perpendicular to magnetic field {lines; hence, the magnetic field angle is found by rotating the polarisation angle by $90$ deg \citep{Davis_Greenstein_1951,Hildebrand_1988,Andersson_et_al_2015}} such that the angle of the field line {is
\begin{equation}
    \theta = \theta^{\prime} + 90^{\circ}.
\end{equation}}

The uncertainty associated with the magnetic field angle is
\begin{equation}\label{polarisation_uncertainty}
    \delta \theta = \frac{1}{2} \times \frac{\sqrt{Q^2 \delta U^2 + U^2\delta Q^2}}{Q^2 + U^2} \times \frac{180^{\circ}}{\pi}.
\end{equation}

The polarisation fraction represents the {fractional amount of polarised emission from the measured total emission}. It is given by
\begin{equation}\label{eq:polarisation_fraction}
    P = \frac{\sqrt{Q^2 + U^2}}{I},
\end{equation}
{where $I$ is the total intensity (Stokes $I$)}. The sum in quadrature of Stokes \textit{Q} and \textit{U} parameters means that the polarisation fraction is always positive. The uncertainty in the polarisation fraction is
\begin{equation}\label{eq:polarisation_fraction_uncertainty}
    \delta P = \frac{\sqrt{Q^2 \delta Q^2 + U^2 \delta U^2 + P^4 I^2 \delta I^2 }}{P I^2},
\end{equation}
where $PI$ is the polarised intensity and $\delta P$ is calculated at each pixel.

\subsection{The Ophiuchus Molecular Cloud}

The Ophiuchus molecular cloud is a nearby, well-studied region of low-to-intermediate-mass star formation \citep{wilking2008}, located $\sim 140$\,pc from the Sun \citep{ortizleon2018}.  The region is made up of two central dark clouds, L1688 and L1689 \citep{Lynds_1962}.  In this work, we consider six column density peaks within Ophiuchus: three in L1688 ($\rho$ Oph A, B and C), and three in L1689 (IRAS 16293, L1689A and L1689B).

\subsubsection{$\rho$ Oph A}

$\rho$ Oph A is a site of active star formation, hosting a number of dense cores.  The submillimetre dust emission is dominated by the Class 0 protostar VLA 1623 \citep{andre1993}, and the dense core SM1 \citep{wardthompson1989}, which is gravitationally bound and likely contains an extremely young embedded source \citep{friesen2014}.  The region is under stellar feedback from two B stars, HD 147889 and S1 \citep[e.g.][]{wilking2008}.

\subsubsection{$\rho$ Oph B}

$\rho$ Oph B is also a site of active star formation, with a complex internal geometry and a number of dense cores and protostars \citep[e.g.][]{pattle2015}, including the outflow-driving protostar IRS 47 \citep[e.g.][]{white2015}.

\subsubsection{$\rho$ Oph C}

$\rho$ Oph C is a quiescent region which contains one or a few starless cores \citep[e.g.][]{pattle2015}, and which is not currently undergoing active star formation.  

\subsubsection{IRAS 16293-2422}

The north of L1689 contains the well-studied protostellar system IRAS 16293-2422, a multiple system of Class 0 protostars \citep{wootten1989,mundy1992}.  The region also contains the starless core IRAS 16293E, which is a candidate for gravitational collapse \citep{sadavoy2010a}.  We refer to this area collectively as `IRAS 16293'.

\subsubsection{L1689A}

L1689A (also known as SMM-16; \citealt{nutter2006}) is a strong candidate for being a gravitationally bound prestellar clump or core \citep{chitsazzadeh2014}.  \citet{pattle2015} and \citet{ladjelate2020} both identified three fragments within L1689A, suggesting that the region is a starless clump rather than a single core. 

\subsubsection{L1689B}

The L1689B prestellar core candidate is embedded in a filamentary structure to the east of the main body of L1689 \citep{jessop2000,kirk2007,steinacker2016}.  The core, which is generally considered to be undergoing large-scale infall \citep[e.g.][]{lee2001}, has been extensively studied due to its relative isolation and simple morphology.

\section{Results} \label{sec:results}

\subsection{Histogram of Relative Orientation}
\label{sec:HRO_description}
An HRO is a comparative tool used to characterise relative alignment between column density and magnetic field structure \citep{Soler_et_al_2013}. It categorises the relative alignment between column density gradients and magnetic field lines as either preferentially parallel or perpendicular.

{The orientation angle of column density iso-density contours is \citep{Soler_et_al_2013}
\begin{equation}\label{eq:iso_contour_angle}
    \psi^{\prime} \equiv \text{arctan} \left( \frac{\partial N / \partial x}{\partial N/ \partial y} \right),
\end{equation}
for projected column density gradients $\partial N / \partial x$ and $\partial N / \partial y$ in the $x-$ and $y-$ directions, respectively, {which correspond to negative Right Ascension and positive Declination}. The overall column density gradient at each pixel is perpendicular to the iso-contour. Its angle, $\psi$,  is given by $\psi = \psi^{\prime} + 90^{\circ}$.}

The angle of relative orientation, $\phi$, is the magnitude of the difference between the angles $\psi$ and $\theta$ as given in Section~\ref{sec: theta_phi_definitions}. In terms of electric fields, we have
\begin{equation}
\phi = \tan^{-1}\left(\frac{|\nabla N\times \hat{E}|}{\nabla N\cdot \hat{E}}\right)
\end{equation}
where $\hat{E}$ is the electric field direction, which is perpendicular to the magnetic field {orientation}. 

The magnetic field constitutes a pseudo-vector, since its absolute direction is unknown. We therefore use the smaller of the two angles created between the column density gradient and magnetic field. All angles of relative orientation are therefore given as angles between zero and 90 degrees.

\subsection{Shape Parameter} \label{sec:shape_param_description}
The shape parameter quantifies the relative alignment between column density gradient and magnetic field line as a function of column density. From an initial HRO, we define two quantities; $A_{c}$ and $A_{e}$. These are defined as in \citet{soler_et_al_2017}, with $A_{c}$ being the area of the histogram where $\phi < 22.5^{\circ}$ and $A_{e}$ the area where $\phi > 67.5^{\circ}$. The shape parameter is then given by \citep{Planck_et_al_2016,soler_et_al_2017,Micelotta_et_al_2021}
\begin{equation}\label{eq:shape_parameter}
    \xi = \frac{A_{c}-A_{e}}{A_{c}+A_{e}},
\end{equation}
where normalisation restricts the value of $\xi$ to between $-1$ and $+1$. $\xi < 0$ indicates that the magnetic field is preferentially perpendicular to the iso-density contours and parallel to the density gradient.  $\xi > 0$ indicates that the magnetic field is preferentially parallel to the iso-density contours, and perpendicular to the density gradient. If $\xi\approx0$, then there is no apparent preferred orientation.

The variances in the central and extreme region values $A_{c}$ and $A_{e}$ are given by $\sigma_{c}^2$ and $\sigma_{e}^2$, respectively. The overall uncertainty in the shape parameter, $\sigma_{\xi}$, is then given by 
\begin{equation}\label{eq:shape_parameter_uncertainty}
    \sigma_{\xi} = \sqrt{ \frac{2 \left( A_{e}^{2} \sigma_{c}^{2} + A_{c}^{2} \sigma_{e}^{2} + A_{c}^{2} \sigma_{c}^{2} + A_{e}^{2} \sigma_{e}^{2} \right) }{ (A_{c}+A_{e})^{4}} }.
\end{equation}

We calculate the shape parameter within each column density bin, which are then plotted as functions of column density. Also plotted is a least-squares regression best-fit line with gradient $C_{\text{HRO}}$ and transition column density exponent $X_{\text{HRO}}$ \citep{soler_et_al_2017}. The equation of the best-fit line is then given by
\begin{equation}\label{chro_xhro_best_fit_equation}
    \xi = C_{\text{HRO}} \left[\text{log}_{10} \left(N({\rm H}_{2}) / \text{cm}^{-2} \right) - X_{\text{HRO}}  \right],
\end{equation}
such that $\xi = 0$ at $N({\rm H}_{2}) = 10^{X_{\text{HRO}}}$.

\subsection{HROs in models} \label{sec:model_hros}
{We created a simple toy model of a spheroidal dense core threaded by (a) linear and (b) hourglass magnetic fields.  We modelled the core as having a Plummer column density distribution \citep{Plummer_1911}, contained within a pixel grid measuring 16,501 pixels wide and 16,501 pixels high. The density distribution was defined as in \citet{Myers_et_al_2020}, assuming an oblate spheroid geometry, with its symmetry axis parallel to the magnetic field axis and zero inclination angle with respect to the plane of the sky. We chose to determine the model in terms of column {density, rather than volume density}, for ease of comparison with observational data. The column density decreases with increasing radius as
\begin{equation}\label{eq:plummer_core_definition}
    \Delta N = \frac{\pi N_{0} r_{0} A}{\sqrt{1+(\xi^{\prime}/A)^{2} + \zeta^{2}}},
\end{equation}
for central column density $N_{0}$,  characteristic core size $r_{0}$, core aspect ratio $A$ and normalised $x-$ and $z-$ coordinates $\xi^{\prime} = x/r_0$ and $\zeta = z/r_0$. Note that $\xi^{\prime}$ is so named to distinguish it from the HRO parameter, $\xi$. The column density at each point was then given as the sum of $\Delta N$ and background column density, $N_{u}$. A background of $N_{u}=10^{20}$ cm$^{-2}$ was used, along with a central column density of $N_{0}=10^{24}$ cm$^{-2}$. This allowed us to study our model core over approximately the same centre-to-edge column density range seen in real cores \citep[e.g.][]{ladjelate2020}. We considered a range of core eccentricities ($e=0, 0.5, 0.8$), for which $A=\sqrt{1-e^{2}}$.} For the linear magnetic field case, we considered cases where the field is orientated parallel to the minor axis of the core, 30 deg offset, 60 deg offset, and perpendicular to the minor axis of the core. For the hourglass magnetic field case, we consider only the case where the hourglass is oriented along the minor axis of the core \citep[see, e.g.,][]{Mouschovias_1976}.

\subsubsection{Linear field model}  \label{sec:linear_models}
The linear magnetic field was defined in each pixel from model Stokes \textit{I}, \textit{Q} and \textit{U} parameter values. In all cases, we defined a polarisation fraction of 10\%.  For our model, an orientation of $0^{\circ}$ corresponds to the linear field being parallel to the core's minor axis, while a core orientation of $90^{\circ}$ places the field perpendicular to the core minor axis.

For our zero eccentricity core model in Fig.~\ref{fig: no_noise_models} we see that all rotation angles correspond to the same shape parameter value of $\xi = 0$ when there is a linear magnetic field. This is expected due to rotational symmetry. 

\subsubsection{Hourglass field model}
{We also created an hourglass magnetic field structure to compare with the linear cases, using the model described by \citet{Myers_et_al_2020}, assuming no toroidal twisting of the magnetic field {and assuming that the hourglass field is entirely in the plane of the sky}. The angle of the magnetic field relative to the symmetry axis at each pixel was defined as
\begin{equation}\label{eq:hourglass_definition}
    \theta_{B} = \text{arctan}\left( \frac{1-t}{s^{-1} + st A^{-2}} \right),
\end{equation}
where $s=\xi^{\prime} / \zeta$ and $t$ is the ratio of density to mean density, defined following equation 5 of \citet{Myers_et_al_2020},  such that 
\begin{equation}
    t = \frac{1 + \nu_{0}(1+\omega^{2})^{-1}}{1 + 3\nu_{0}\omega^{-2}(1-\omega^{-1}\tan^{-1}\omega)}
\end{equation}
where
\begin{equation}
  \omega = \sqrt{\left(\frac{\xi^{\prime}}{A}\right)^{2}+\zeta^{2}} 
\end{equation}
and the centre-to-edge volume density contrast $\nu_{0}$ is given by
\begin{equation}
    \nu_{0} = \frac{\Delta N_{max}}{N_{u}r_{0}A}.
\end{equation}
Note that $\theta_{B}$ is a function of aspect ratio $A$, and so the shape of the hourglass varies with the eccentricity of the core.

The dimensions of the pixel grid allowed us to define the pixel at the origin as the central pixel. The hourglass field was aligned with the minor axis of the core (see Appendix~\ref{sec:appendix_model_figs} for model visualisations).

Data selection from the model was based on column density contours. The minimum column density studied for each model was that which corresponded to the largest complete density contour surrounding the core. It was found in each case that the minimum closed contour of column density within the defined pixel grid was $10^{21.8}$ cm$^{-2}$. Of the $16501\times16501$ pixels, a total of $203,680,473$ ($\sim75\%$) were kept for $e=0$, decreasing to $176,392,445$ ($\sim65\%$) for $e=0.5$ and $122,208,329$ ($\sim45\%$) for $e=0.8$.} We thus generated a set of magnetic field pseudo-vectors which were then compared to the column density gradient vectors of our model core.

The model gives us a chance to understand what an HRO would look like in these situations. The model acts as a prediction for the results expected from performing an HRO analysis on real cores. 

In Fig.~\ref{fig: no_noise_models} we plot the shape parameter profiles in our initial model setup for cores with eccentricity 0.0, 0.5 and 0.8, respectively. In each case we see the profile obtained for an hourglass field (black), along with linear magnetic field lines parallel to the model core's minor axis (green), then at angle of 30 deg (blue), 60 deg (purple) and perpendicular (red). Each show the shape parameters as a function of column density. {In each case 25 column density bins were used. All of the HROs for hourglass fields show flat profiles at low column density, tending towards that of the parallel linear field. The profiles then diverge to a minimum in $\xi$ at medium column density. At the highest densities the hourglass field models converge back to the parallel linear-field values of $\xi$.}

\begin{figure}
    \centering
        \includegraphics[width=0.99\columnwidth]{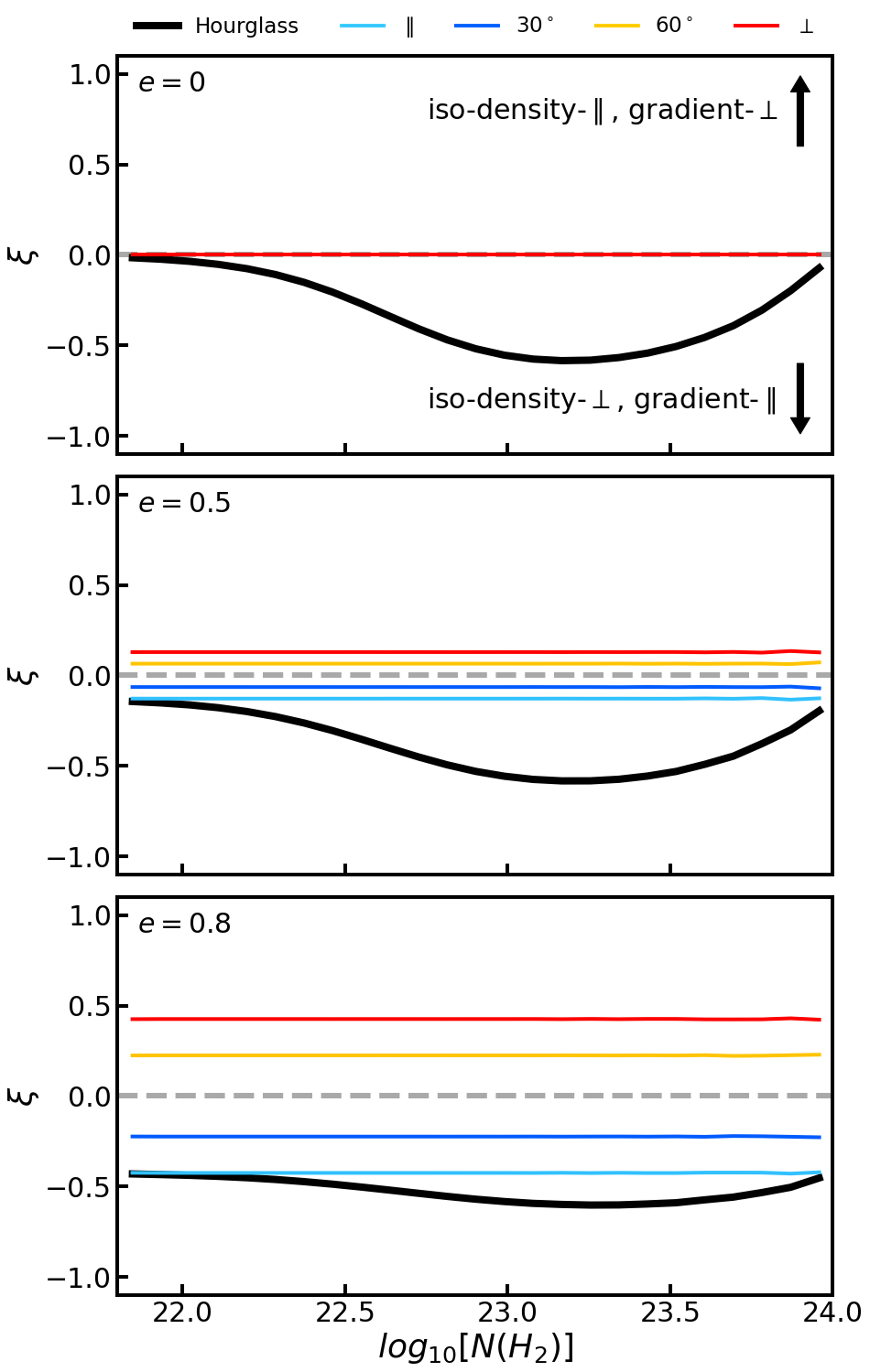}
        \caption{HRO shape parameter $\xi$ for a model cores of eccentricity 0 (i.e. circular, top), 0.5 (middle) and 0.8 (bottom) for different magnetic field structures with no noise. Profiles for an hourglass field and {multiple} linear magnetic fields at varying rotation angles from the minor axis -- parallel to axis, $30^{\circ}$, $60^{\circ}$, and perpendicular -- are all shown.}
        \label{fig: no_noise_models}
\end{figure}

\begin{figure}
    \centering
        \includegraphics[width=0.99\columnwidth]{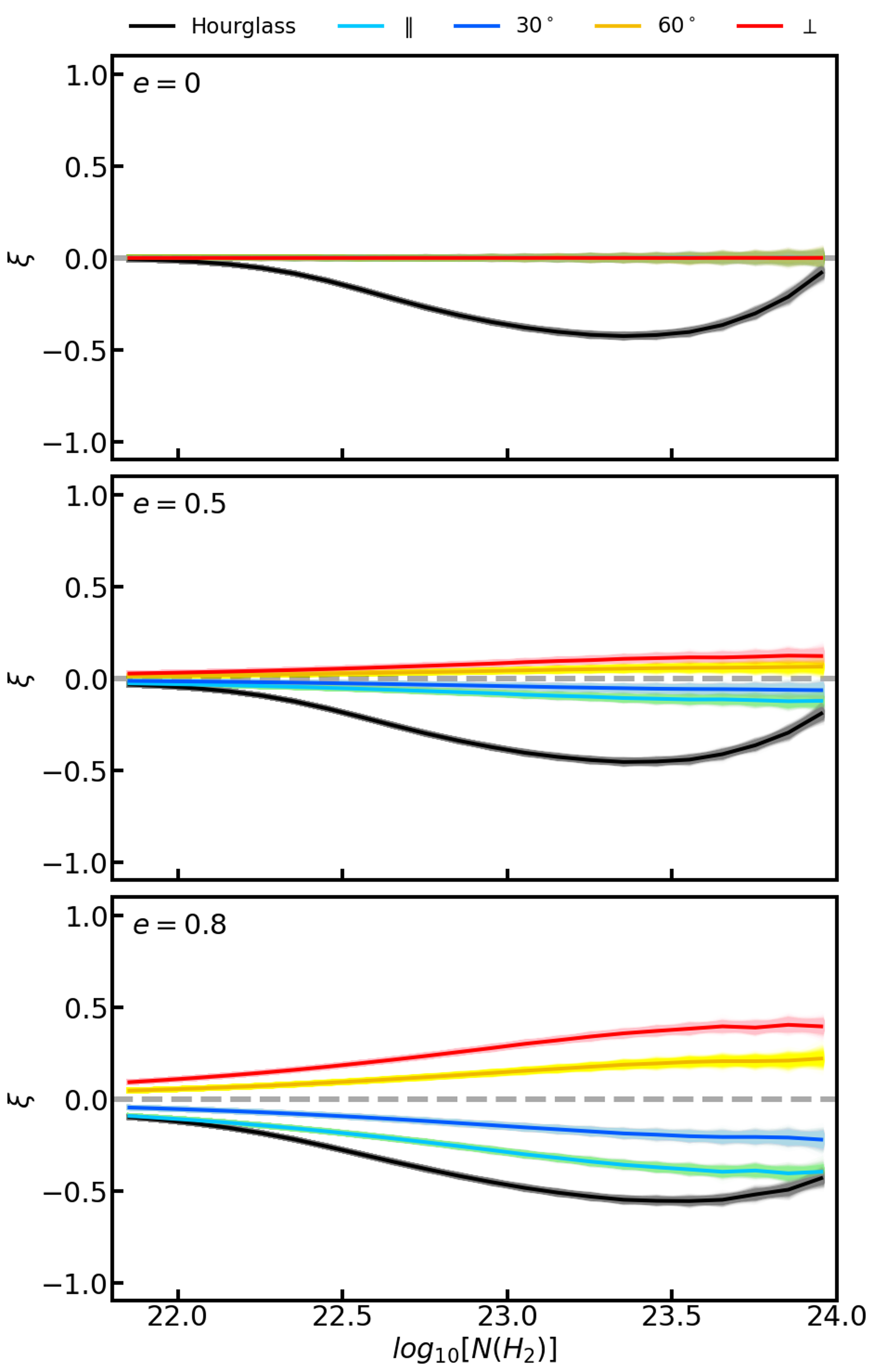}
        \caption{HRO shape parameter $\xi$ for a model cores of eccentricity 0 (i.e. circular, top), 0.5 (middle) and 0.8 (bottom) for different magnetic field structures with added noise. Profiles for an hourglass field and {multiple} linear magnetic fields at varying rotation angles from the minor axis -- parallel to axis, $30^{\circ}$, $60^{\circ}$, and perpendicular -- are all shown.}
        \label{fig: noise_models}
\end{figure}

{\subsection{Modelling noise} \label{sec:added_noise}
{In order to investigate the effects that including instrumental noise has on the models, we added noise at each pixel. }

{In each case, the magnetic field pseudo-vectors were decomposed into Stokes $I$, $Q$ and $U$ parameters. Molecular clouds are expected to show a power-law relationship between polarisation fraction and total intensity, such that $p \propto I^{-\alpha}$ \citep{Whittet_et_al_2008}. We adopted a polarisation model in which polarisation fraction decreases with increasing brightness (thus column density), given by the power law
\begin{equation}
    p = p_{0} \left( \frac{I}{I_{0}}\right)^{-\alpha}
\end{equation}
with index $\alpha$, where $I_{0}$ is the minimum intensity value in our pixel grid and $p_{0}$ is the polarisation fraction at the edge of the grid. We determined $I_{0}$ for each grid individually on each run and assumed 10\% polarisation at the edge of the pixel grid, i.e. $p_{0}=0.1$. Observations of Oph A show an index of $\alpha\approx0.34$, while for Oph B and C they show $\alpha \approx 0.6-0.7$ \citep{pattle2019}. We therefore selected the intermediate value of $\alpha=0.5$ for all noise models. The resulting polarised intensity, 
\begin{equation}
    PI = p \times I
\end{equation}
was then used to find the decomposed Stokes $Q$ and $U$ parameters
\begin{equation}
    Q^{\prime} = \frac{PI}{\sqrt{1+\tan^{2}{(2\theta_{B})}}}
\end{equation}
and
\begin{equation}
    U^{\prime} = PI \times \sin{(2\theta_{B})}
\end{equation}
where $\theta_{B}$ is given in equation~(\ref{eq:hourglass_definition}). Both $Q^{\prime}$ and $U^{\prime}$ are initially noise-free.}

{The noise value at each pixel was randomly drawn from a Gaussian distribution centred on zero, with a width of $2\times 10^{21}$ cm$^{-2}$, then added to $Q^{\prime}$ and $U^{\prime}$. In our models, we have a signal-to-noise ratio in Stokes $I$ of $\sim 3$ at the edge of the map and $\sim 500$ at the centre.}

{We see that adding noise to our models results in the loss of a measurable preferred alignment at low column densities, as shown in Fig.~\ref{fig: noise_models}. In these cases, the behaviour that we observe transitions from $\xi \approx 0$ at low column densities, to the value of $\xi$ expected for the core in question. High column densities are almost entirely unaffected. Noisy HROs becomes indistinguishable from our noiseless models when the Stokes $I$ SNR $\gtrsim 6$, after which the HROs are identical. This occurs at a column density of $1.7\times10^{23}$ cm$^{-2}$.}

\subsection{HROs in Ophiuchus} \label{sec:oph_hros}
To compare the model with observation, we performed an HRO analysis on Ophiuchus. In regions of low signal-to-noise ratios, statistical biasing acts to artificially increase the polarisation fraction \citep[e.g.][]{Vaillancourt_2006}. In order to substantially reduce the bias at low signal-to-noise ratios, we calculated debiased polarisation fractions using \citep{Wardle_Kronberg_1974}
\begin{equation}\label{eq:debiassed_polarisation_fraction}
   P_{\text{db}} = \frac{\sqrt{Q^2 + U^2 - \frac{1}{2}\left( \delta Q^2 + \delta U^2 \right)}}{I},
\end{equation}

where $\delta Q$ and $\delta U$ are the uncertainties in Stokes \textit{Q} and \textit{U} values. 

We selected data based on three criteria:{ $\text{SNR} \geq 3$, $\text{SNR}_{I}\geq3$, and $\delta \theta \leq 10\text{ deg}$,} where $\text{SNR} = P_{\text{db}} / \delta P$ and $\text{SNR}_{I} = I / \delta I$, for {total intensity} uncertainty $\delta I$ and magnetic field angle uncertainty $\delta \theta$.  We note that the first and third of these criteria should be degenerate \citep{serkowski1962}.

Fig.~\ref{fig:obs_hros} shows $\xi$ values as a function of column density for the regions that we consider.  We see that regions of Ophiuchus exhibit different trends in $\xi$ with increasing column density using \textit{Herschel} and JCMT data. Ophiuchus, L1689 and $\rho$ Oph A show flat or very slightly increasing values of $\xi$, but with generally negative (thus parallel to the density gradient) values. L1688, $\rho$ Oph B, $\rho$ Oph C, IRAS 16293 and L1689A show an increase in $\xi$ that is steeper, along with a transition from negative to positive (perpendicular) values. The transition column densities for L1688 and IRAS 16293 are extrapolated values based on the best-fit line, while the transitions for $\rho$ Oph B, $\rho$ Oph C and L1689A are observed within the data. Finally, L1689B is the only region to show a negative gradient in $\xi$ and a transition from positive-$\xi$ (perpendicular) to negative (parallel), {although it is important to note that this trend is based on only two data points}.

{In order to confirm the statistical accuracy of the HRO profile, we performed bootstrapping on the data. The Stokes $Q$ and $U$ maps were individually re-sampled 1000 times. For each pixel on the grid, the standard deviation of the values of the pixels surrounding it was used to generate a Gaussian of the same width. A new value was then randomly chosen from within this Gaussian. Fig.~\ref{fig:obs_hros} demonstrates that all re-sampled HRO profiles fall within the uncertainties calculated using equation~(\ref{eq:shape_parameter_uncertainty}). In the cases where equation~(\ref{eq:shape_parameter_uncertainty}) estimates very large uncertainties, we find the re-sampled HROs show considerably less variation. Overall, we find the trends in the HRO shape factor are statistically significant.}

\begin{figure*}
    \centering
    \includegraphics[width=\textwidth]{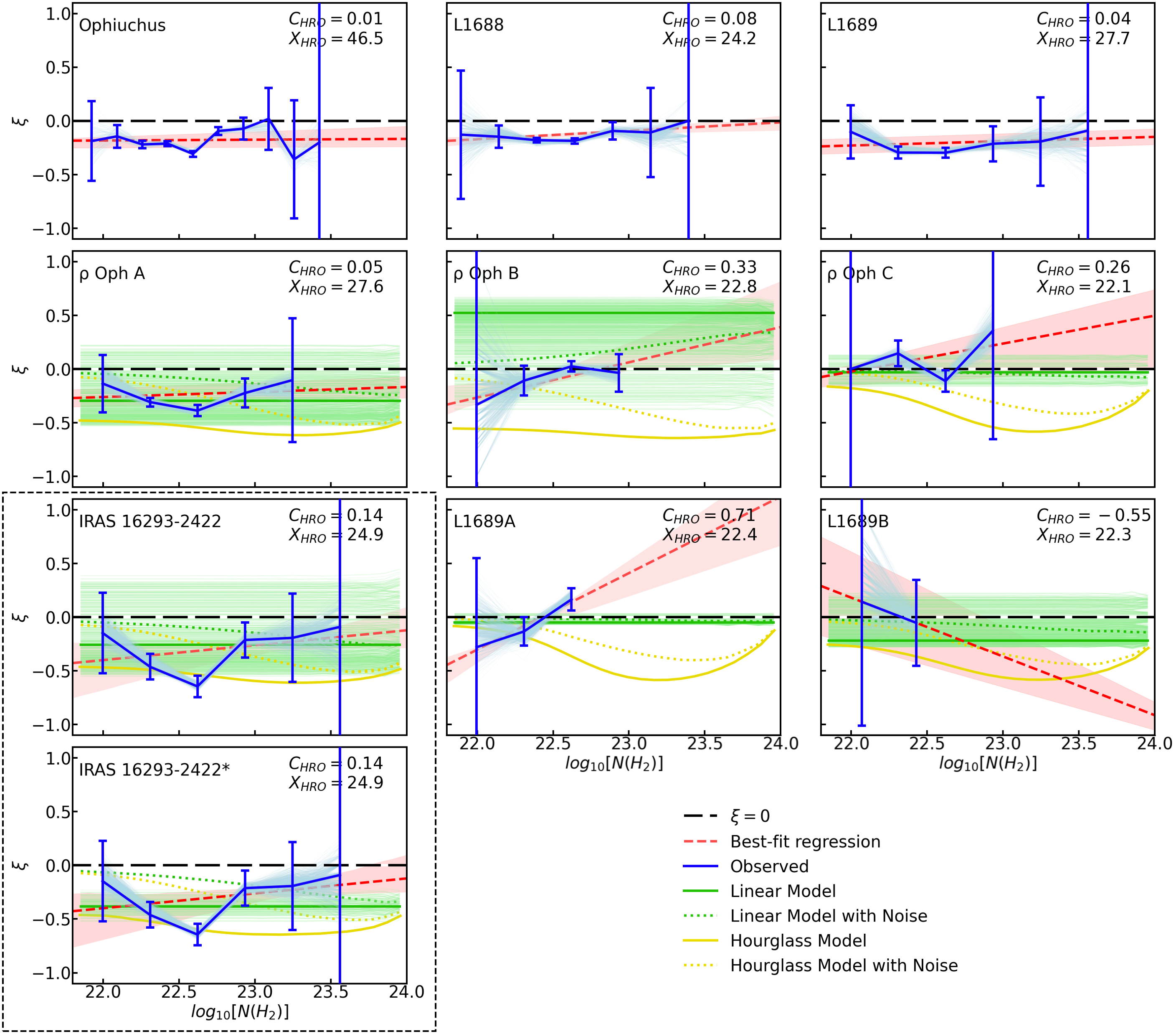}
    \caption{{HRO shape parameter $\xi$ (blue) and associated uncertainty (equations~(\ref{eq:shape_parameter}) and~(\ref{eq:shape_parameter_uncertainty})) determined for hydrogen column density bins in each region of Ophiuchus. We also show HROs produced from re-sampling the observational data (light blue) and the least-squares regression best-fit line (dashed red) {for the observational HROs, along with the} corresponding values of $C_{\text{HRO}}$ and $X_{\text{HRO}}$. From $\rho$ Oph A to L1689B, we also show the HRO for a linear magnetic field calculated from a Gaussian fit on the regions (green) and the uncertainty (light green) based on re-sampling. We further show the HRO for an hourglass magnetic field (yellow). The region IRAS 16293-2422* shows a second HRO calculated for IRAS 16293 from a second data cut (see Section~\ref{sec:gaussian_fitting}). The black dashed line represents the $\xi = 0$ line. {The HRO for L1688 used the combined data from $\rho$ Oph A, B and C as shown in Fig.~\ref{fig:L1688_gaussians}. Similarly, the HRO for L1689 used the combined data from IRAS 16293-2422, L1689A and L1689B in Fig.~\ref{fig:L1689_gaussians}. The HRO for Ophiuchus then used the combined data from L1688 and L1689.}}}
    \label{fig:obs_hros}
\end{figure*}

\begin{figure}
    \includegraphics[width=0.96\columnwidth]{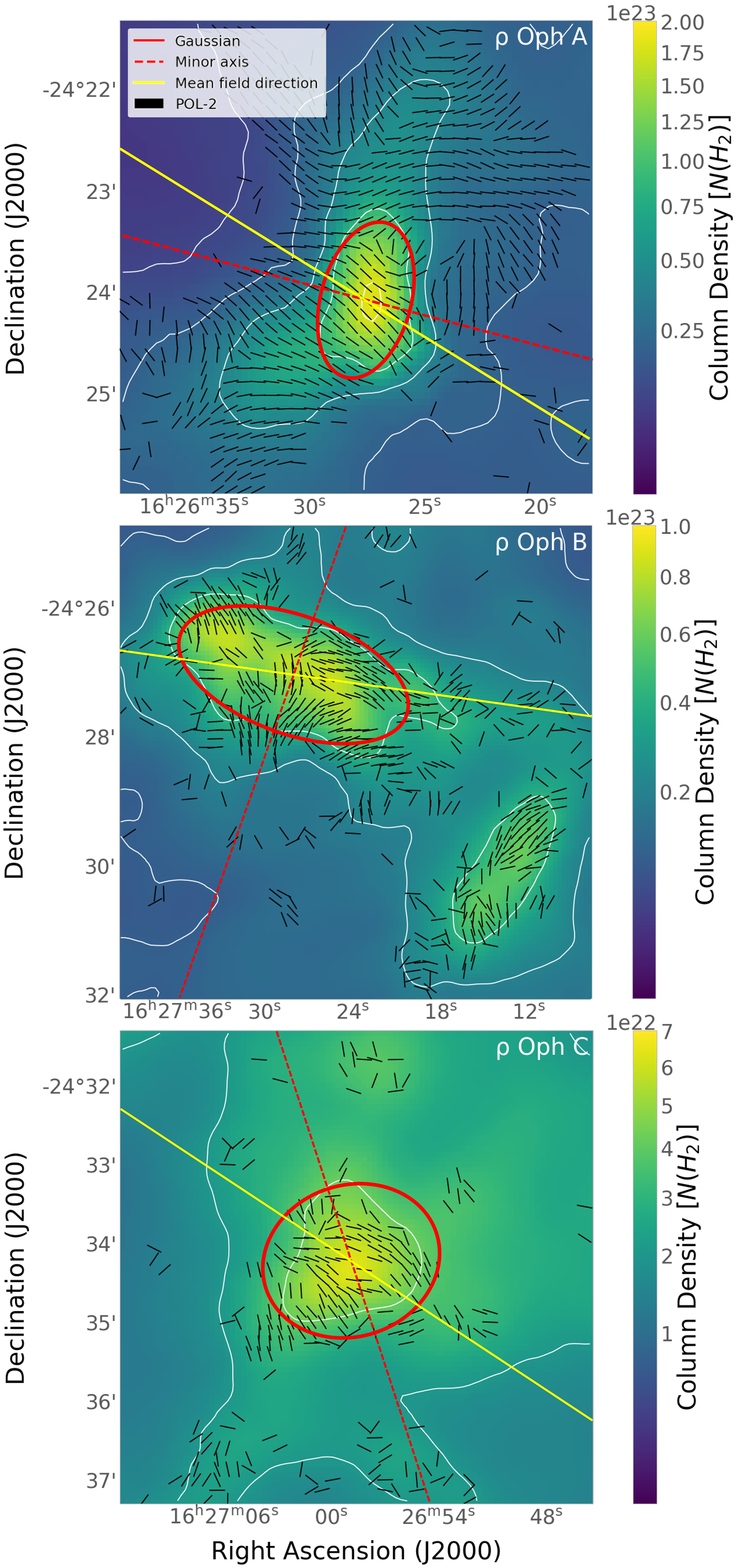}
    \caption{The regions $\rho$ Oph A (top), $\rho$ Oph B (middle) and $\rho$ Oph C (bottom) within the Ophiuchus cloud L1688. Background data show the \textit{Herschel} column density map for each region. The POL-2 magnetic field pseudo-vectors are overlaid in black, with the Gaussian fit for each region shown in solid red and its minor axis in dashed red. The solid yellow line represents the mean magnetic field direction.}
    \label{fig:L1688_gaussians}
\end{figure}

\begin{figure}
    \includegraphics[width=0.96\columnwidth]{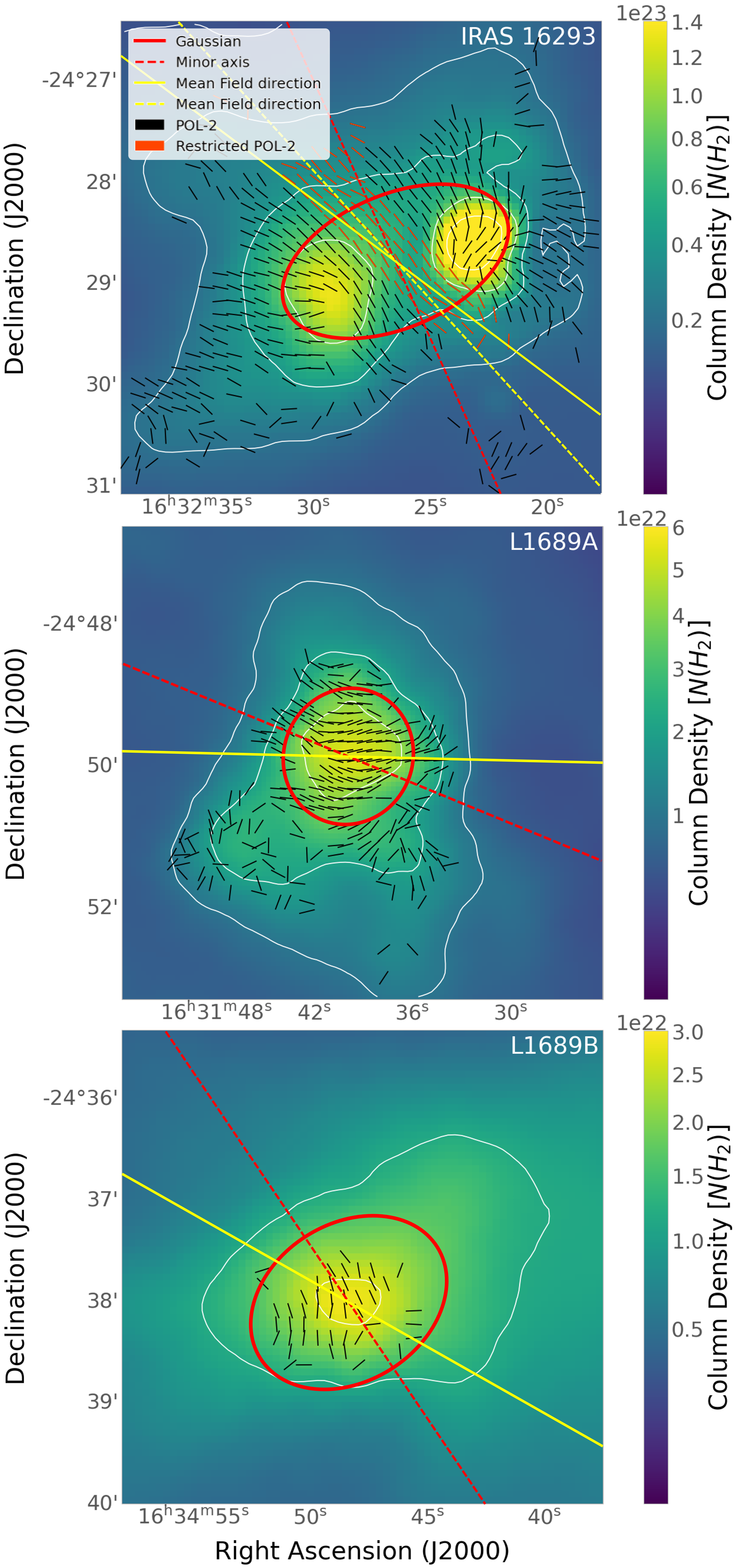}
    \caption{The regions IRAS 16293 (top), L1689A (middle) and L1689B (bottom) within the Ophiuchus cloud L1689. Background data show the \textit{Herschel} column density map for each region. The POL-2 magnetic field line pseudo-vectors are overlaid in black, with the Gaussian fit for each region shown in solid red and its minor axis in dashed red. The solid yellow line represents the mean magnetic field direction. For IRAS 16293, an additional data cut limited magnetic field vectors to those in orange. Their mean direction is shown in dashed yellow.}
    \label{fig:L1689_gaussians}
\end{figure}

\subsection{Gaussian fitting} \label{sec:gaussian_fitting}
In order to test the predictive capabilities of our toy model, we studied the magnetic field morphology of each region of Ophiuchus. We fitted two-dimensional Gaussian distributions, as in \citet{Pattle_et_al_2021}, but to the column density maps within each region rather than Stokes $I$ maps, in order to estimate their sizes, eccentricities and position angles. The fitted Gaussians were created using all $N({\rm H}_2)$ data satisfying the signal-to-noise and uncertainty cuts in Section~\ref{sec:oph_hros} and produced ellipses with specified central positions, sizes, major axis rotation angles and eccentricities. The {Gaussian fits} for each region are shown in red in Figs.~\ref{fig:L1688_gaussians} and~\ref{fig:L1689_gaussians}. We then calculated the mean magnetic field direction for each region, based only on the magnetic field pseudo-vectors contained within the fitted Gaussian ellipse. The vectors were chosen with the same criteria as in Section~\ref{sec:shape_param_description}, and their mean angle was taken to be the mean magnetic field {orientation} in the core.

We calculated the relative angle between the Gaussian fit minor axis and the mean magnetic field {orientation}. The relative angle and core eccentricity values are shown in Table~\ref{tab:gaussian_fits}. These values were used as input values into our model, such that we were able to create tailored models for each core. In each case, the magnetic field was defined as a linear field at a known relative angle. The resultant HRO shape parameter profiles are shown as solid blue lines in Fig.~\ref{fig:obs_hros}. {We again performed bootstrapping on our Gaussian models, completing 1000 runs for each model. In each case, random values for Gaussian eccentricity and position angle and mean B-field angle were selected from within the uncertainty ranges shown in Table~\ref{tab:gaussian_fits}. The resulting HRO profiles are plotted individually in Fig.~\ref{fig:obs_hros}. Given that we modelled a completely linear magnetic field, the possible shape parameter profiles are all horizontal lines within the uncertainty region. We repeated this process for our hourglass field, with the Gaussian fit HRO shown in yellow in Fig.~\ref{fig:obs_hros} along with bootstrapping uncertainties. We then added noise to the Gaussian HRO setups and repeated the process.}

In contrast to \citet{Pattle_et_al_2021}, the Gaussian fitting for IRAS 16293 is not dominated by emission from the IRAS 16293-2422 protostellar system when using \textit{Herschel} column density maps. The use of a data mask was therefore not required. We see in Fig.~\ref{fig:L1689_gaussians}, however, that there is both large-scale order to the magnetic field in IRAS 16293, and significant ordered deviation around the IRAS 16293 protostars and the IRAS 16293E core. In order to model only the large-scale magnetic field, we performed an additional data cut after the initial SNR and $\delta \theta$ cuts and excluded magnetic field vectors close to the most dense regions in IRAS 16293. The vectors included are shown in Fig.~\ref{fig:L1689_gaussians}. The resultant best-fit Gaussian data is shown in Table~\ref{tab:gaussian_fits} and the corresponding shape parameter profile in Fig.~\ref{fig:obs_hros}, {both under the name under IRAS 16293*. We see that the associated uncertainty} is much smaller than when no additional data cut is applied.

\begin{table*}
	\centering
	\caption{Gaussian fits to column density maps from the regions within Ophiuchus.  {Note:} IRAS 16293* is the Gaussian fit after a data cut that excludes observations close to the high-density regions of IRAS 16293 as described in Section~\ref{sec:gaussian_fitting}.}
	\label{tab:gaussian_fits}
	\begin{tabular}{lccccccc} 
		\hline
		   & $\rho$ Oph A & $\rho$ Oph B & $\rho$ Oph C & IRAS 16293 & IRAS 16293* & L1689A & L1689B\\
		\hline
		Centre R.A. (hh:mm:ss.ss) & 16:26:27.65 & 16:27:28.31 & 16:26:59.07 & 16:32:26.52 & 16:32:26.52& 16:31:40.00& 16:34:48.35\\
		Centre decl. (dd:mm:ss.s) & $-24$:24:02.1 & $-24$:26:59.2 & $-24$:34:09.9 & $-24$:28:46.2 & $-24$:28:46.2 & $-24$:49:51.4 & $-24$:38:01.0\\
            $a$ (major std. dev.) (arcsec) & $45\pm2$ & $105\pm22$ & $65\pm3$ & $64\pm7$ & $64\pm7$ & $69\pm2$ & $62\pm3$\\
            $b$ (minor std. dev.) (arcsec) & $22\pm4$ & $50\pm21$ & $55\pm11$ & $36\pm9$ & $36\pm9$ & $64\pm2$ & $47\pm5$\\
		Eccentricity & $0.83\pm0.03$ & $0.87\pm0.05$ & $0.53\pm0.03$ & $0.82\pm0.05$ & $0.82\pm0.05$ & $0.37\pm0.01$ & $0.65\pm0.04$\\
            Position Angle (deg E of N) & $85\pm10$ & $-19\pm1$ & $18\pm2$ & $24\pm1$ & $24\pm1$ & $67\pm34$ & $34\pm5$\\
            Mean Field Angle (deg E of N) & $58\pm28$ & $82\pm50$ & $57\pm38$ & $53\pm38$ & $42\pm10$ & $89\pm22$ & $47\pm59$\\
            Relative Angle (deg) & $27\pm30$ & $79\pm50$ & $39\pm38$ & $29\pm38$ & $18\pm10$ & $22\pm40$ & $13\pm59$\\
            \hline
	\end{tabular}
        \label{tab:L1688_gaussian_fits}
\end{table*}

\begin{table}
	\centering
	\caption{HRO shape parameter best-fit line equations for each region shown in Fig.~\ref{fig:obs_hros}, such that {$\xi = C_{\text{HRO}} \left[\text{log}_{10} \left(N({\rm H}_{2}) / \text{cm}^{-2} \right) - X_{\text{HRO}}  \right]$}.}
	\label{tab:best_fits}
	\begin{tabular}{ccc} 
		\hline
    Region & {$C_{\rm HRO}$}& {$X_{\rm HRO}$}\\
		\hline
            Ophiuchus & $+0.01^{+0.04}_{-0.01}$ & {$46.5^{+25.5}_{-21.5}$} \\[5pt]
            L1688 & $+0.08\pm0.05$ & {$24.2^{+3.4}_{-0.6}$} \\[5pt]
            L1689 & $+0.04\pm0.01$ & {$27.7^{+1.8}_{-2.0}$} \\[5pt]
            $\rho$ Oph A & $+0.05\pm0.01$ & {$27.6^{+3.5}_{-2.3}$} \\[5pt]
            $\rho$ Oph B & $+0.33\pm0.23$ & {$22.8^{+0.4}_{-0.2}$} \\[5pt]
            $\rho$ Oph C & $+0.26^{+0.09}_{-0.18}$ & {$22.1^{+1.8}_{0.2}$} \\[5pt]
            IRAS 16293 & $+0.14^{+0.24}_{-0.14}$ & {$24.9^{+125}_{-1.1}$} \\[5pt]
            L1689A & $+0.71\pm0.24$ & {$22.4\pm0.1$} \\[5pt]
            L1689B & $-0.55\pm0.15$ & {$22.3^{+0.6}_{-0.9}$} \\[5pt]
		\hline
	\end{tabular}
\end{table}

\section{Discussion} \label{sec:discussion}
In this section, we discuss the results of an HRO analysis of our toy model. We discuss the results of our initial modelling, before discussing how well the regions in Ophiuchus compare to the best-fit models.

\subsection{Model Overview}
We find that in the absence of noise, every linear model produces a constant shape parameter.  For position angles $< 45$ degrees, $\xi < 0$, while for position angles $> 45$ degrees, $\xi > 0$.  The greatest offsets from $\xi = 0$ are for the parallel and perpendicular cases, and the magnitude of this offset increases with core eccentricity.  For $e=0$, $\xi = 0$ for all position angles.

This HRO behaviour can be simply understood as resulting from the elongation of highly eccentric cores relative to the magnetic field direction, as illustrated in Fig.~\ref{fig:core_diagram}.  Where a core has high eccentricity and a low position angle, a large fraction of core has column density gradient close to parallel to magnetic field, so $\xi < 0$.  However, the ends of the ellipse will always have a gradient perpendicular to the magnetic field direction, so $\xi > -1$.  Conversely, for cores with a high eccentricity and a high position angle, a large fraction of the core has a column density gradient close to perpendicular to the magnetic field, so $\xi > 0$.  However, the ends of the ellipse will always have a gradient parallel to the magnetic field direction, so $\xi < 1$.

For the case where $e = 0$, there is no preferred orientation between magnetic field direction and column density, so $\xi = 0$.  

A preferentially parallel geometry (with respect to column density gradient) is the sign of a core elongated perpendicular to the magnetic field direction -- as is predicted for gravitational collapse in a strong magnetic field -- and the magnetically-dominated hourglass models studied here produce a parallel geometry. The column density gradients of our models cores point towards the centre of the mass because the cores have a monotonically increasing density distribution. 

{The hourglass geometry in every case produces an HRO which tends towards the parallel linear model at low column densities. At intermediate column densities the hourglass HRO diverges from that of the linear model, falling to a minimum value of $\xi$. It does this until the highest column densities, at which point it converges back on the parallel-linear-field behaviour ($-1 \leq \xi(e) < 0$).  By using the hourglass field defined by \citet{Myers_et_al_2020}, we find that the large pixel grid used in our modelling is sufficient in size to allow the most distant field lines to have almost zero curvature. We are using the shape parameter as defined in Section~\ref{sec:shape_param_description}, so to be classed as parallel, the magnetic field lines need only be oriented within 22.5 deg of the column density gradients. At low column densities, the hourglass field mimics the linear field. For intermediate densities the curvature of the magnetic field is such that a much larger proportion of field lines are within 22.5 deg of the column density gradients, causing the increase to more negative values of $\xi$. At high column densities, the hourglass geometry produces a field that is again very similar to the linear model.}

The difference between the linear value of $\xi$ and the minimum value in an hourglass magnetic field is dependent on the eccentricity of the core, with smaller eccentricities having a larger separations. The minimum value itself is consistent for all eccentricities, taking a value of $\sim -0.6$, varying very slightly with decreasing $e$.

{The addition of noise allows us to understand the observed HRO profiles in Fig.~\ref{fig:obs_hros}. When re-sampling of Stokes $Q$ and $U$ maps is performed, the original shape parameter values fall at the centre of the distribution of re-sampled values. A similar result is found when applying the same technique to our models in Fig.~\ref{fig: noise_models}}.

{We also see the observational data have larger uncertainties at smaller column densities than at intermediate. In our toy models, the largest effect of adding noise was to alter the appearance of the HRO at low column densities, increasing the uncertainty in the actual profile.}

{Overall we find that the uncertainty in the observational HROs appears to result from a combination of measurement uncertainty and smaller numbers of data points in the lower and higher column density bins.}

Our toy models suggest the following interpretation for the HROs of spheroidal dense cores: We expect that a core with a geometry corresponding to strongly magnetised star formation (moderate eccentricity, with an angle $\lesssim 30$ deg between the core minor axis and the magnetic field direction) will have a preferentially parallel HRO ($-1 < \xi < 0$), with no transition between perpendicular and parallel behaviours, but possibly a transition downward from $\xi \approx 0$ at the lowest column densities.  {Divergence to a minimum value of $\xi \approx -0.6$ is strongly indicative of an hourglass field geometry.}

\subsection{$\rho$ Oph A}
Comparing the observational shape parameter profiles and those modelled on best-fit data in Fig.~\ref{fig:obs_hros}, we see that the model HRO measured from a linear magnetic field at a relative angle determined from our best-fit Gaussian model shows good agreement with the POL-2 data. The parameters determined from a Gaussian fitting enable us to successfully use a toy model to construct a qualitatively accurate HRO for $\rho$ Oph A.

Here, all the observational data points fall within the uncertainty range of the model shape parameter profiles, although some observational uncertainty lies outside. The model shape parameter value stays constant due to the use of a linear magnetic field, yet still does not diverge significantly from the observational best-fit line. {The optimal linear model HRO is almost entirely contained with the 1-sigma uncertainty of the best-fit regression line, which itself shows an almost horizontal profile. }We find that of all regions modelled, $\rho$ Oph A has the best match with our model.

There is a similarity between the behaviour of the shape parameter calculated here with POL-2 data and in \citet{Lee_et_al_2021}, where it is calculated with HAWC+ polarisation data. Their Fig. 3 shows $\xi$ exhibiting significant parallel relative alignment between the magnetic field and (\textit{Herschel}) column density gradient. In particular, over the column density ranges of our work, \citet{Lee_et_al_2021} also find $\xi$ less parallel with increasing column density, moving closer to $\xi=0$. All data points have negative $\xi$. This again is similar to our Fig.~\ref{fig:obs_hros} which uses POL-2 observations. We obtained a strong parallel signal in our HRO, with a best-fit line showing a slight trend towards less parallel alignment with increasing column density. A difference between our results and \citet{Lee_et_al_2021} is seen in the steepness of the trends and the values of $\xi$, but the overall trend is seen in both.

\begin{figure*}
    \centering
    \includegraphics[width=\textwidth]{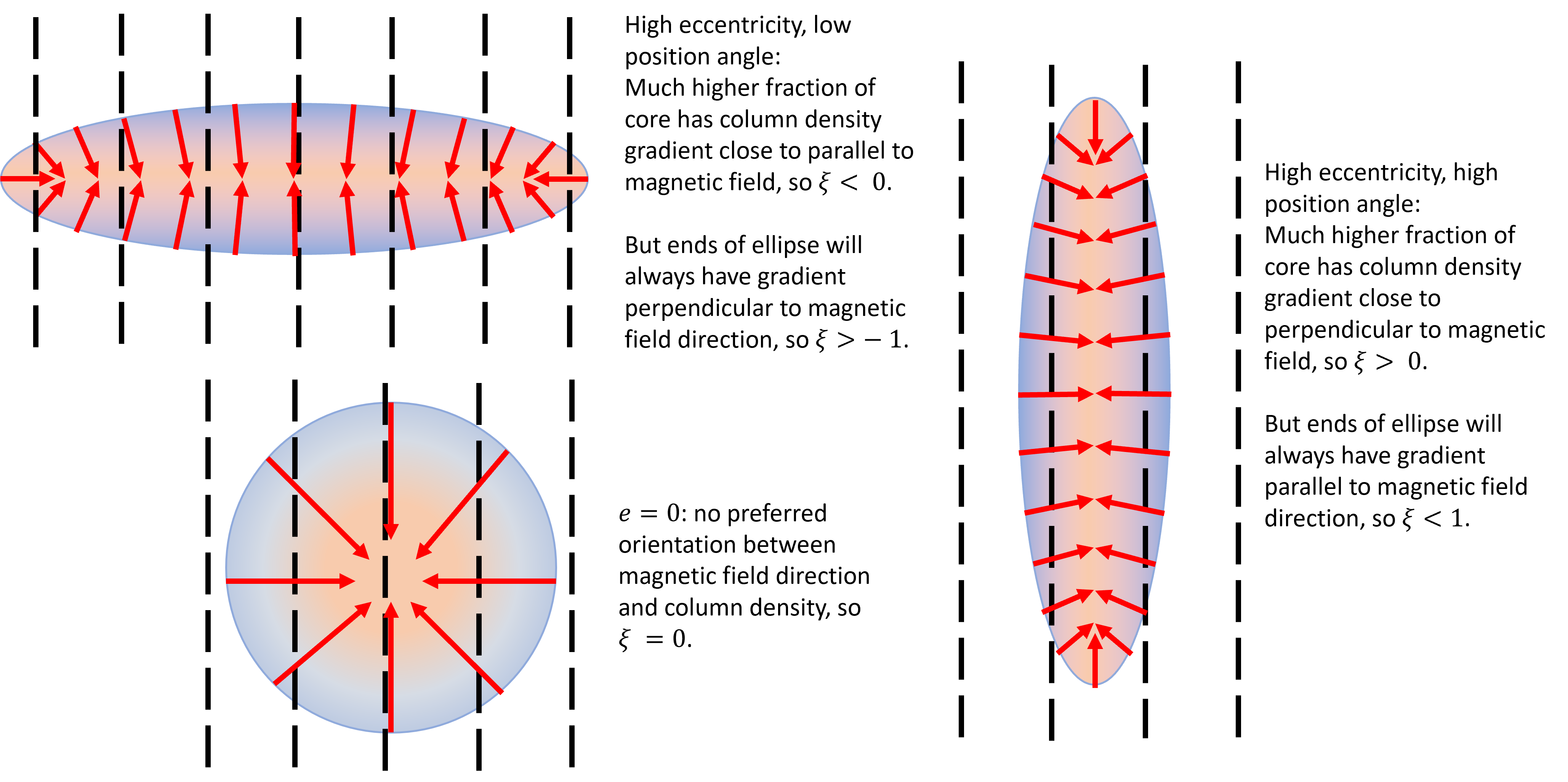}
    \caption{Example core-field geometries with column density gradient vectors (red arrows). Magnetic field lines are shown in black (dashed lines) parallel and perpendicular to cores with high aspect ratio. Also shown is a circular core.}
    \label{fig:core_diagram}
\end{figure*}

A possible reason for the small variations between the datasets is the difference in observing wavelength between HAWC+ and POL-2, with the two observing at 154 and 850 $\mu$m, respectively. The shorter wavelength will preferentially trace dust at a higher temperature.  Oph A is significantly heated by the B-type star S1 (e.g \citealt{pattle2015}, and refs. therein), and so HAWC+ likely mapped polarisation from dust that is warmer, and so nearer the core surface, than did POL-2. Comparing our Fig.~\ref{fig:L1688_gaussians} and Fig. 1 of \citet{Lee_et_al_2021} we see slight variations in magnetic field orientation between the POL-2 and HAWC+ maps. This comparison suggests there may be a genuine difference in behaviour between the two temperature regimes, even though the overall effect on determining the value of the HRO shape parameter is similar.

\subsection{IRAS 16293}
IRAS 16293 observational data are not all contained within the uncertainty of the model profile, which itself is larger than in $\rho$ Oph A. When we consider the additional data cut of restricting study to only linear fields away from the highest-density regions, we find that IRAS 16293 has a much better fit. The uncertainty is significantly reduced, but the shape parameter profile still fits the data well. From this, we find the overall shape parameter in IRAS 16293 is determined mainly from the ordered magnetic field crossing the minor axis in Fig.~\ref{fig:L1689_gaussians}. It also matches Fig.~\ref{fig: no_noise_models}, with the magnetic field being near-parallel to the core minor axis (18 deg). The less well-ordered field near the centres of IRAS 16293 A and B slightly changes the value of $\xi$, but the biggest effect is to increase the uncertainty.

Both $\rho$ Oph A and IRAS 16293 (particularly after the second data cut) are well-behaved starless cores with high aspect ratios with a relatively uniform magnetic field. For these regions, we find the HRO has a strong signal that makes it a good predictor for the behaviour of the magnetic field that we expect to see.

\subsection{$\rho$ Oph B}
The internal structure of $\rho$ Oph B prevented our toy model from achieving a good fit. From Fig.~\ref{fig:obs_hros} we see that the best-fit model HRO shape parameter profile for $\rho$ Oph B is not close to any of the measured HRO data points, and is only within the uncertainty range of one data point.  However, the large uncertainty range on the value of the model shape parameter profile covers most data points from the observational data.  {Despite this large range of $\xi$, we note that there is a strong weighting near the optimal fit Gaussian model HRO at $\xi=0.52$, with $62\%$ of profiles found within 0.2 of the optimal fit. From our model runs, $86\% $ of profiles have $\xi \geq 0$ and $82\%$ have $\xi \geq 0.1$. Only 9\% of profiles have $\left|{\xi}\right|<0.1$. The weighting is such that while the model does not preclude a value of $\xi\sim0$ (which would be consistent with Oph B data), it strongly suggests a value above zero.} This is found to be due to the complex nature of $\rho$ Oph B, which contains considerable substructure, with many embedded cores. Since there is no single region of bright emission that is preventing a good Gaussian fit being made, we did not attempt an additional data cut or masking.  The region is too complex to be accurately modelled as a simple Gaussian column density distribution. We are seeing contributions from all of the substructure in the calculation of both the mean magnetic field angle and the Gaussian fit. 

\subsection{$\rho$ Oph C}
In $\rho$ Oph C, the model shape parameter profile calculated from the best-fit Gaussian parameters does not cross the observational shape parameter profile. In this case, however, we see a closer fit than in $\rho$ Oph B. The fitted profile passes through the uncertainty ranges of three of the four observational data points. The overall uncertainty in the fitted model is also smaller. The magnetic field and column density structure in $\rho$ Oph C are less complex than $\rho$ Oph B. We see a relatively uniform magnetic field which makes our toy model suitable. However, the smaller aspect ratio in $\rho$ Oph C causes a weaker HRO signal, with shape parameter values close to zero.

\subsection{L1689A and L1689B}
Of the regions studied in this work, L1689A and L1689B were determined to have amongst the smallest eccentricities (see Table~\ref{tab:gaussian_fits}). As such, when modelled with a linear magnetic field we expect to see shape parameter values close to zero. This was found to be the case. In particular, we see in Fig.~\ref{fig:obs_hros} that the value of $\xi$ for L1689A is $-0.05$, which comes from its small eccentricity of $e=0.37$. We also note that the uncertainty in the best-fit linear model for L1689A is very small. The model uncertainty is dominated by uncertainties in relative angles between column density gradients and core minor axis. As with $\rho$ Oph C, we have a relatively uniform magnetic field. But the small aspect ratio has negated much of this effect since relative position angles have less meaning as the core becomes more circular. 

\subsection{Predictions from HROs}
Comparing observational HROs with modelled ones demonstrates a further use in predicting the overall shape of the magnetic field of a region. Dense cores with HROs that fall to a minimum in parallel alignment are more likely to have an hourglass magnetic field. Cores with elongated geometries would be expected to have magnetic fields that are approximately linear and parallel to the core's minor axis. The resultant HRO would be expected to be consistently negative, but with $\xi>-1$. We would then expect hourglass magnetic fields to have HROs which match a parallel linear field at low column density, but which diverges to a minimum, before converging back at the highest column density. However, we would expect only cores with significantly flattened geometries to produce strong HRO signals. Those with small aspect ratios would have $\xi\approx0$ no matter what the orientation of the magnetic field.

Our best-fit models consider a linear field and a monotonically increasing Plummer column density distribution. The magnetic field is in a fixed position across all pixels and is not distorted by increasing column density. We see that there are complex processes at work within each region and that those with low aspect ratio and large uncertainty in $\xi$ are more difficult to reproduce with a simple toy model.

\section{Conclusions} \label{sec:conclusion}
We have investigated the expected behaviour of Histograms of Relative Orientation (HROs) in dense spheroidal cores, and applied the results of our analysis to the Ophiuchus molecular cloud. We used JCMT BISTRO Survey polarisation maps and column density maps from the \textit{Herschel} Gould Belt Survey.  We compared the results with HROs created from a simple model containing idealised magnetic field structures -- linear and hourglass -- and Plummer column density distributions. We then fitted two-dimensional Gaussian models to the column density maps for each region of Ophiuchus, using the results as input to our model. The resulting HRO shape parameter profiles were then compared directly with the observational profiles. {The addition of artificial noise in our models enabled us to understand the causes of uncertainty in our observations.}

We find that both $\rho$ Oph A and IRAS 16293 have HROs with consistently negative shape parameters ($\xi$). The column density structures exhibit high aspect ratios, with magnetic fields that are on the whole linear and parallel to the minor axes. $\rho$ Oph C, L1689A and L1689B also contain relatively linear magnetic fields. However, all have low aspect ratios and weak HRO signals, such that all have $\xi$ values of approximately zero in both observation and model. $\rho$ Oph B was too complex to be characterised using our simple model.

We also find that for a well-behaved dense core (i.e. simple structure) with a high aspect ratio the HRO produced from a simple model is a good predictor for the expected behaviour between column density gradients and magnetic field. When no strong aspect ratio is present, then we expect the HRO shape parameter to take a value of $\xi\approx0$. The small eccentricity of such regions means that on their small size scales there is no meaning in defining the magnetic field as parallel or perpendicular to the density structure.

Our model has identified that an idealised starless core formed in a strong magnetic field will have a consistently negative HRO shape parameter.  This is a similar behaviour to that seen on cloud/filament scales. 

A comparison of observational HROs and our model demonstrate that none of the regions studied in Ophiuchus contain an hourglass field. The ability to eliminate this field structure as a possible configuration within a region from only the HRO shape parameter profile is a demonstration of its predictive capabilities.

The analysis conducted here demonstrates that it is possible to study the relationship between column density gradients and magnetic field lines without detailed models, but that applications are limited to well-behaved regions with high aspect ratios. {When noise is added to models we see the largest change of HRO profile is at low column densities. Given that we see large uncertainties in Ophiuchus HRO shape parameter plots at similar values, we find that measurement uncertainty can account for this. We also note that our work necessarily focused on the POS. No three-dimensional information on density or B-field structure was used.}

{In future work we will consider more complex models that include, for example, a second core (where required) and the effects of background structure in the column density map.} Further application of this analysis to larger samples of starless cores is also needed, in order to refine our model (particularly the eccentricity at which an HRO signal becomes clearly detectable), and to identify those most likely to have formed in a strongly magnetised environment.


\section*{Acknowledgements}

K.P. is a Royal Society University Research Fellow, supported by grant no. URF\textbackslash R1\textbackslash 211322.

The James Clerk Maxwell Telescope is operated by the East Asian Observatory on behalf of The National Astronomical Observatory of Japan; Academia Sinica Institute of Astronomy and Astrophysics; the Korea Astronomy and Space Science Institute; the National Astronomical Research Institute of Thailand; Center for Astronomical Mega-Science (as well as the National Key R\&D Program of China with No. 2017YFA0402700). Additional funding support is provided by the Science and Technology Facilities Council of the United Kingdom and participating universities and organizations in the United Kingdom, Canada and Ireland.  Additional funds for the construction of SCUBA-2 were provided by the Canada Foundation for Innovation.

This research made use of data from the \textit{Herschel} Gould Belt survey (HGBS) project (\url{http://gouldbelt-Herschel.cea.fr}). The HGBS is a \textit{Herschel} Key Programme jointly carried out by SPIRE Specialist Astronomy Group 3 (SAG 3), scientists of several institutes in the PACS Consortium (CEA Saclay, INAF-IFSI Rome and INAF-Arcetri, KU Leuven, MPIA Heidelberg), and scientists of the \textit{Herschel} Science Center (HSC).

This research has made use of: Starlink software \citep{currie2014}, currently supported by the East Asian Observatory; Astropy (\url{http://www.astropy.org}), a community-developed core Python package and an ecosystem of tools and resources for astronomy \citep{astropy:2013, astropy:2018, astropy:2022}; NASA's Astrophysics Data System.

The authors wish to recognize and acknowledge the very significant cultural role and reverence that the summit of Maunakea has always had within the indigenous Hawaiian community.  We are most fortunate to have the opportunity to conduct observations from this mountain.

\section*{Data Availability}

The raw data used in this paper are available in the JCMT archive under project codes M16AL004 and M19AP038.  Maps and vector catalogues for L1689 are available from \url{https://doi.org/10.11570/20.0013}.  Maps and vector catalogues for L1688 are available at \url{http://star.ucl.ac.uk/BISTRO/}.



\bibliographystyle{mnras}




\appendix

\section{Models} \label{sec:appendix_model_figs}

Below we show visualisations of the model cores used in this paper. We show our initial models as described in Section~\ref{sec:model_hros}. 

\begin{figure}
    \centering
        \includegraphics[width=0.75\columnwidth]{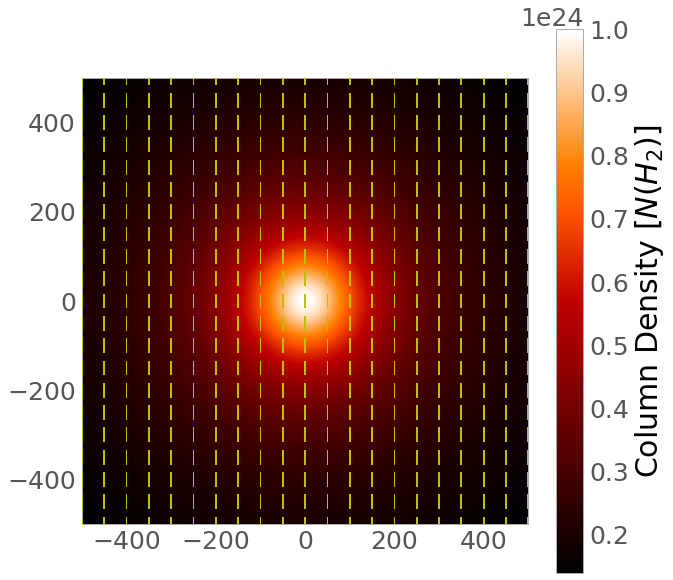}
        \caption{Model core with eccentricity $e=0$, with vertical magnetic field line (dashed yellow lines).}
        \label{fig: mod_e00_mag_para}
    \medskip

        \includegraphics[width=0.75\columnwidth]{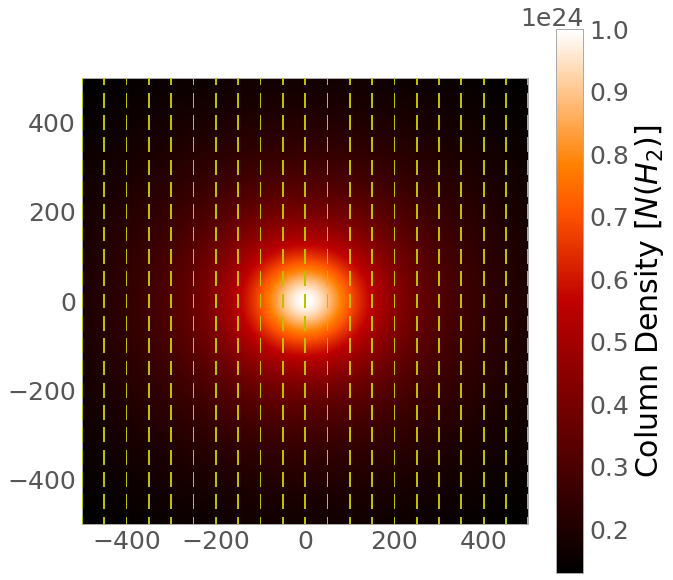}
        \caption{Model core with eccentricity $e=0.5$, rotated such that its minor axis is parallel to a linear vertical magnetic field line (dashed yellow lines).}
        \label{fig: mod_e05_mag_para}
    \medskip

        \includegraphics[width=0.75\columnwidth]{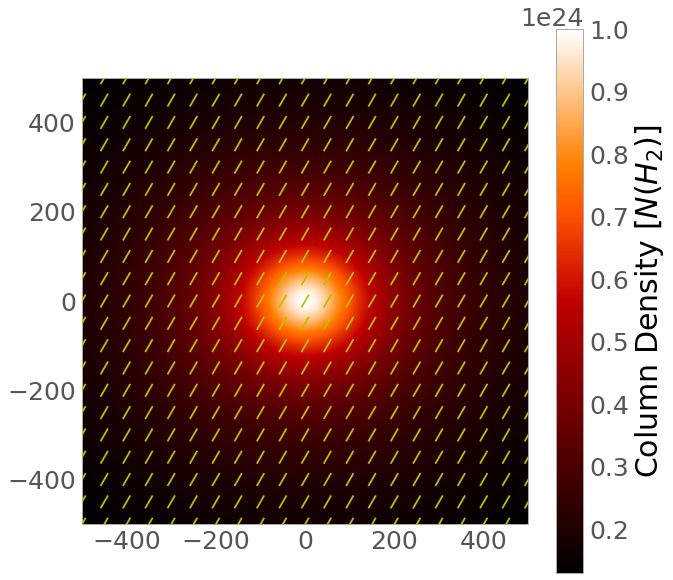}
        \caption{Model core with eccentricity $e=0.5$, rotated such that its minor axis is at an angle of $30^{\circ}$ to a linear vertical magnetic field line (dashed yellow lines).}
        \label{fig: mod_e05_mag_30}
\end{figure}

\begin{figure}
    \centering
        \includegraphics[width=0.75\columnwidth]{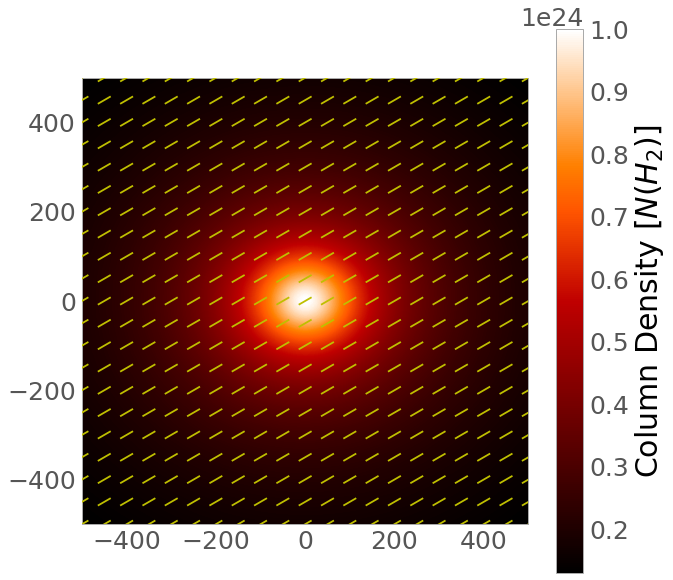}
        \caption{Model core with eccentricity $e=0.5$, rotated such that its minor axis is at an angle of $60^{\circ}$ to a linear vertical magnetic field line (dashed yellow lines).}
        \label{fig: mod_e05_mag_60}
    \medskip

        \includegraphics[width=0.75\columnwidth]{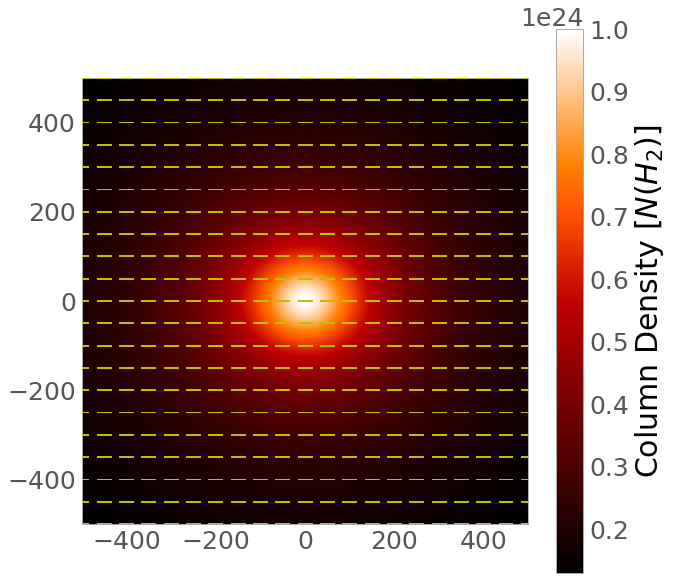}
        \caption{Model core with eccentricity $e=0.5$, rotated such that its minor axis is perpendicular to a linear vertical magnetic field line (dashed yellow lines).}
        \label{fig: mod_e05_mag_perp}
    \medskip

        \includegraphics[width=0.75\columnwidth]{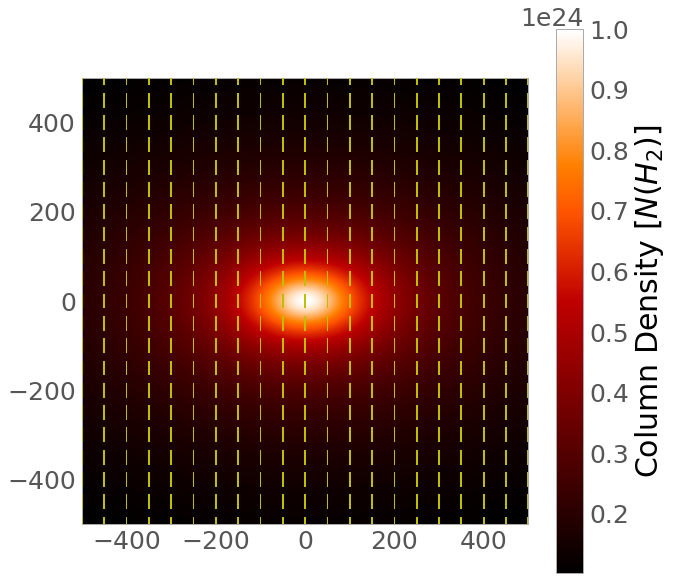}
        \caption{Model core with eccentricity $e=0.8$, rotated such that its minor axis is parallel to a linear vertical magnetic field line (dashed yellow lines).}
        \label{fig: mod_e08_mag_para}
\end{figure}
    
\begin{figure}
    \centering    
        \includegraphics[width=0.75\columnwidth]{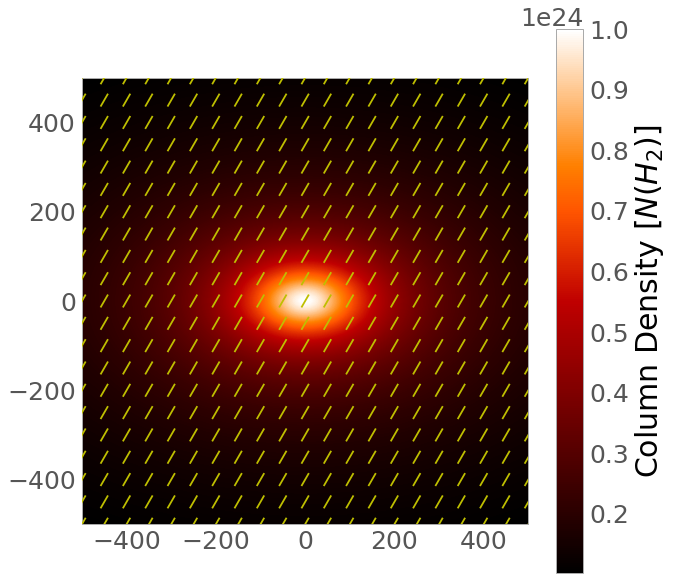}
        \caption{Model core with eccentricity $e=0.8$, rotated such that its minor axis is at an angle of $30^{\circ}$ to a linear vertical magnetic field line (dashed yellow lines).}
        \label{fig: mod_e08_mag_30}
    \medskip

        \includegraphics[width=0.75\columnwidth]{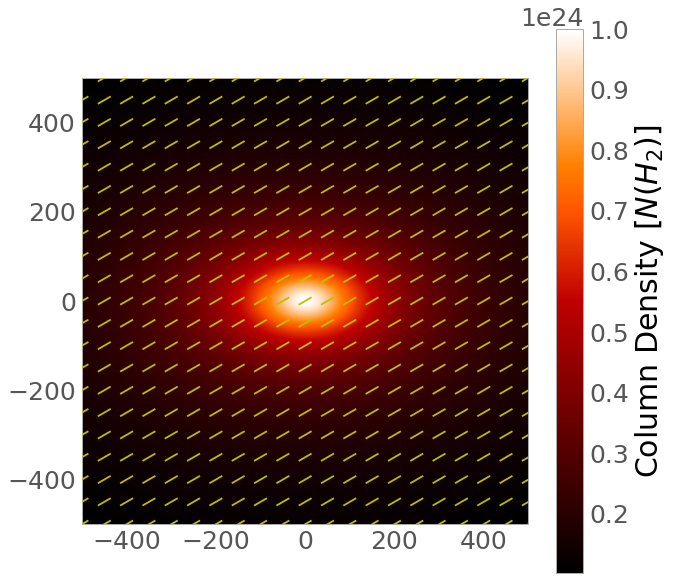}
        \caption{Model core with eccentricity $e=0.8$, rotated such that its minor axis is at an angle of $60^{\circ}$ to a linear vertical magnetic field line (dashed yellow lines).}
        \label{fig: mod_e08_mag_60}
    \medskip
    
        \includegraphics[width=0.75\columnwidth]{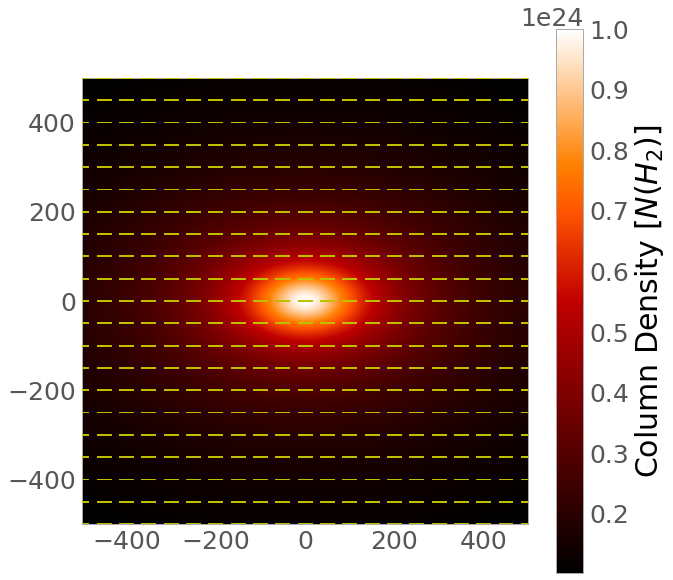}
        \caption{Model core with eccentricity $e=0.8$, rotated such that its minor axis is perpendicular to a linear vertical magnetic field line (dashed yellow lines).}
        \label{fig: mod_e08_mag_perp}
\end{figure}

\begin{figure}
    \centering    
        \includegraphics[width=0.75\columnwidth]{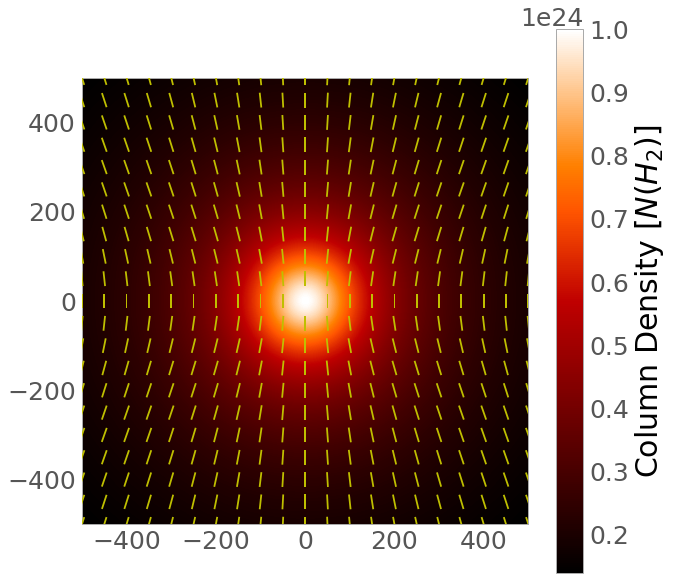}
        \caption{Model core with eccentricity $e=0$ within an hourglass magnetic field (dashed yellow lines).}
        \label{fig: hourglass_model_e00}
    \medskip

        \includegraphics[width=0.75\columnwidth]{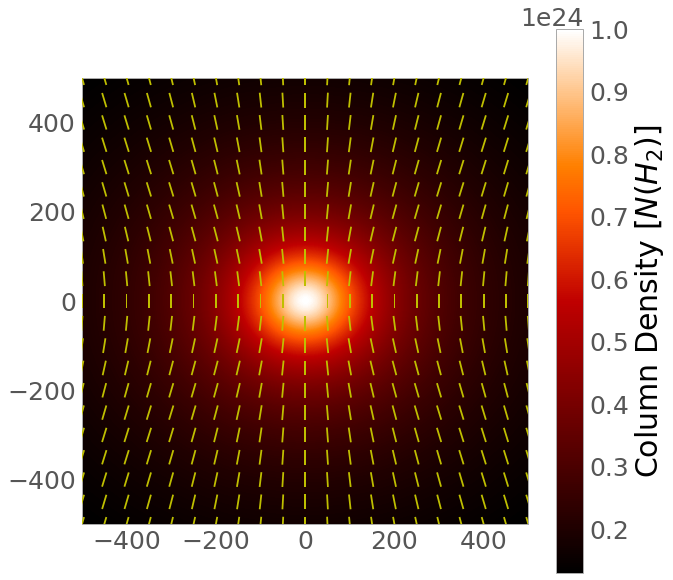}
        \caption{Model core with eccentricity $e=0.5$, rotated such that its minor axis is perpendicular to the middle of an hourglass magnetic field (dashed yellow lines).}
        \label{fig: hourglass_model_e05}
    \medskip
    
        \includegraphics[width=0.75\columnwidth]{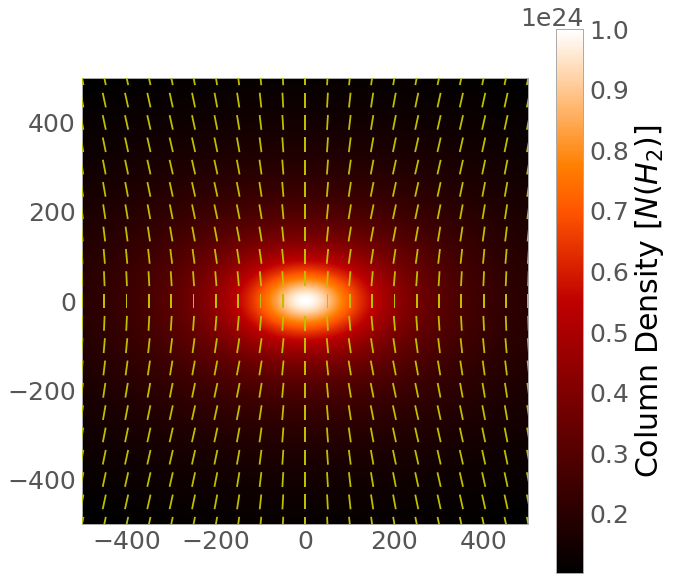}
        \caption{Model core with eccentricity $e=0.8$, rotated such that its minor axis is perpendicular to the middle of an hourglass magnetic field (dashed yellow lines).}
        \label{fig: hourglass_model_e08}
\end{figure}

\section{Effects of noise} \label{sec:appendix_example_noise}

Below we show a map of magnetic field pseudo-vectors for our $e=0.5$ hourglass model when noise is absent and present. We see that the effect of the noise diminishes close to the centre of the grid. The core itself has been removed from Fig.~\ref{fig: appendix_vector_comparison}.

\begin{figure*}
    \centering
    \includegraphics[width=\textwidth]{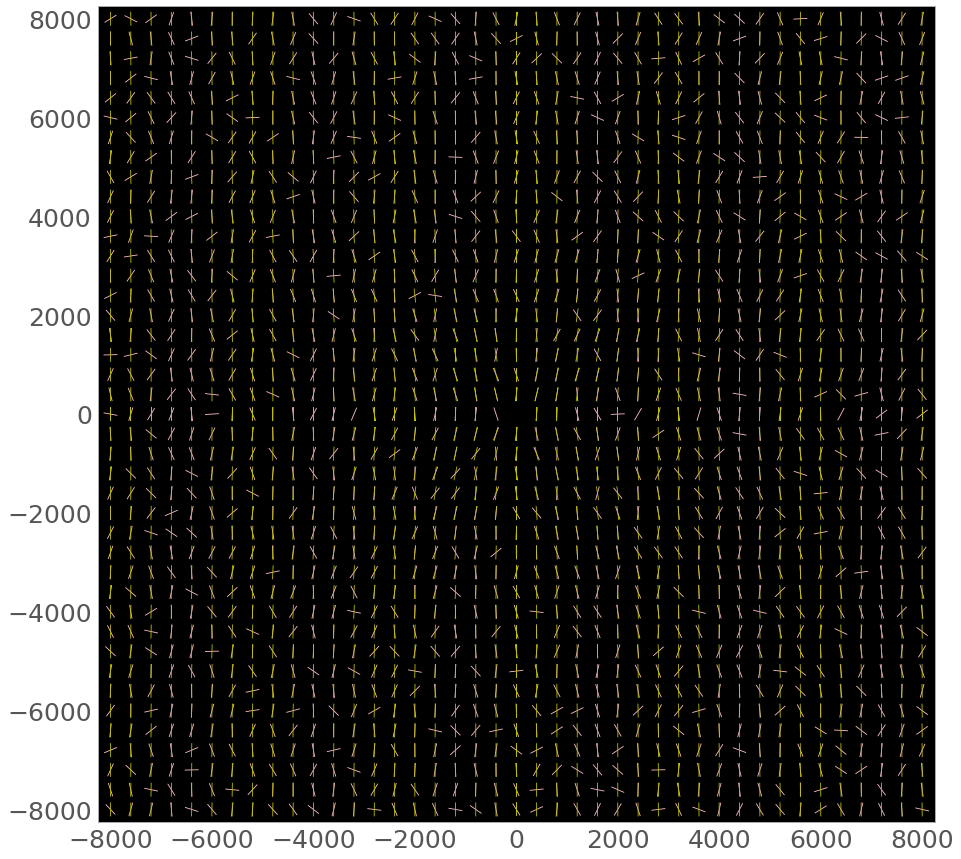}
    \caption{Comparison of $e=0.5$ hourglass model magnetic field pseudo-vectors both with (yellow) and without (pink) noise.}
    \label{fig: appendix_vector_comparison}
\end{figure*}
    

\bsp	
\label{lastpage}
\end{document}